\PassOptionsToPackage{unicode}{hyperref}
\PassOptionsToPackage{hyphens}{url}
\PassOptionsToPackage{dvipsnames,svgnames,x11names}{xcolor}
\documentclass[
]{article}

\usepackage{amsmath,amssymb}
\usepackage{iftex}
\ifPDFTeX
  \usepackage[T1]{fontenc}
  \usepackage[utf8]{inputenc}
  \usepackage{textcomp} 
\else 
  \usepackage{unicode-math}
  \defaultfontfeatures{Scale=MatchLowercase}
  \defaultfontfeatures[\rmfamily]{Ligatures=TeX,Scale=1}
\fi
\usepackage{lmodern}
\ifPDFTeX\else  
\fi
\IfFileExists{upquote.sty}{\usepackage{upquote}}{}
\IfFileExists{microtype.sty}{
  \usepackage[]{microtype}
  \UseMicrotypeSet[protrusion]{basicmath} 
}{}
\makeatletter
\@ifundefined{KOMAClassName}{
  \IfFileExists{parskip.sty}{%
    \usepackage{parskip}
  }{
    \setlength{\parindent}{0pt}
    \setlength{\parskip}{6pt plus 2pt minus 1pt}}
}{
  \KOMAoptions{parskip=half}}
\makeatother
\usepackage{xcolor}
\ifx\paragraph\undefined\else
  \let\oldparagraph\paragraph
  \renewcommand{\paragraph}[1]{\oldparagraph{#1}\mbox{}}
\fi
\ifx\subparagraph\undefined\else
  \let\oldsubparagraph\subparagraph
  \renewcommand{\subparagraph}[1]{\oldsubparagraph{#1}\mbox{}}
\fi

\usepackage{longtable,booktabs,array}
\usepackage{calc} 
\usepackage{etoolbox}
\makeatletter
\patchcmd\longtable{\par}{\if@noskipsec\mbox{}\fi\par}{}{}
\makeatother
\IfFileExists{footnotehyper.sty}{\usepackage{footnotehyper}}{\usepackage{footnote}}
\makesavenoteenv{longtable}
\usepackage{graphicx}
\makeatletter
\def\maxwidth{\ifdim\Gin@nat@width>\linewidth\linewidth\else\Gin@nat@width\fi}
\def\maxheight{\ifdim\Gin@nat@height>\textheight\textheight\else\Gin@nat@height\fi}
\makeatother
\setkeys{Gin}{width=\maxwidth,height=\maxheight,keepaspectratio}
\makeatletter
\def\fps@figure{htbp}
\makeatother

\usepackage{arxiv}
\usepackage{orcidlink}
\usepackage{amsmath}
\usepackage[T1]{fontenc}
\usepackage[utf8]{inputenc}
\usepackage{pifont}
\usepackage{newunicodechar}
\usepackage{amssymb}
\newunicodechar{✓}{\ding{51}}
\newunicodechar{✗}{\ding{55}}
\newunicodechar{̆}{\u{}}
\newunicodechar{∈}{\in}
\makeatletter
\@ifpackageloaded{caption}{}{\usepackage{caption}}
\AtBeginDocument{%
\ifdefined\contentsname
  \renewcommand*\contentsname{Table of contents}
\else
  \newcommand\contentsname{Table of contents}
\fi
\ifdefined\listfigurename
  \renewcommand*\listfigurename{List of Figures}
\else
  \newcommand\listfigurename{List of Figures}
\fi
\ifdefined\listtablename
  \renewcommand*\listtablename{List of Tables}
\else
  \newcommand\listtablename{List of Tables}
\fi
\ifdefined\figurename
  \renewcommand*\figurename{Figure}
\else
  \newcommand\figurename{Figure}
\fi
\ifdefined\tablename
  \renewcommand*\tablename{Table}
\else
  \newcommand\tablename{Table}
\fi
}
\@ifpackageloaded{float}{}{\usepackage{float}}
\floatstyle{ruled}
\@ifundefined{c@chapter}{\newfloat{codelisting}{h}{lop}}{\newfloat{codelisting}{h}{lop}[chapter]}
\floatname{codelisting}{Listing}

\makeatother
\makeatletter
\@ifpackageloaded{caption}{}{\usepackage{caption}}
\@ifpackageloaded{subcaption}{}{\usepackage{subcaption}}
\makeatother
\ifLuaTeX
  \usepackage{selnolig}  
\fi
\usepackage{bookmark}

\IfFileExists{xurl.sty}{\usepackage{xurl}}{} 
\hypersetup{
  pdftitle={What is Escalation? Measuring Crisis Dynamics in International Relations with Human and LLM Generated Event Data},
  pdfauthor={, , , , and },
  colorlinks=true,
  linkcolor={blue},
  filecolor={Maroon},
  citecolor={Blue},
  urlcolor={Blue},
  pdfcreator={LaTeX via pandoc}}

\renewcommand{\today}{2024-01-17}
\newcommand{\runninghead}{A Preprint }
\title{What is Escalation? Measuring Crisis Dynamics in International
Relations with Human and LLM Generated Event Data}
\def\asep{\\\\\\ } 
\def\asep{\And }
\author{\textbf{Rex W. Douglass}\\\\University of California, San
Diego\\\\\asep\textbf{Erik Gartzke}\\\\University of California, San
Diego\\\\\asep\textbf{Jon R. Lindsay}\\\\Georgia Institute of
Technology\\\\\asep\textbf{J. Andrés Gannon}\\\\Vanderbilt
University\\\\\asep\textbf{Thomas Leo Scherer}\\\\University of
California, San Diego\\\\}
\date{2024-01-17}
\begin{document}
\maketitle
\begin{abstract}
When a dangerous international crisis begins, leaders need to know
whether their next move is going to resolve the dispute or amplify it
out of control. Theories of conflict have mainly served to deepen the
confusion, revealing fighting, bargaining, and signaling to be
high-dimensional and subtle equilibrium behaviors with deeply contextual
consequences. Should a leader communicate resolve through aggressive
acts, avoid spirals through accommodation, or focus on ensuring the
possibility of a bargain? We offer a data-driven empirical solution to
this logjam in the form of a new large-scale analysis of actions taken
within 475 crises. We combine two complimentary measurement projects,
the human-coded International Crisis Behavior Events (ICBe) dataset and
the new machine-coded ICBeLLM. We model directly whether an action tends
to shorten or extend the length of a crisis. The result is a directly
interpretable measure of the latent escalatory/deescalatory nature of
each action leaders have chosen over the last century.
\end{abstract}

\textsuperscript{1} University of California, San Diego\\
\textsuperscript{2} Georgia Institute of Technology\\
\textsuperscript{3} Vanderbilt University

\section{Introduction}\label{introduction}

What does it mean that an international crisis is escalating, and what
actions within a crisis are escalatory? This question haunts recent and
ongoing crises such as missile tests on the Korean Peninsula, Russian
incursions in Ukraine, partial state death in Syria, and Chinese claims
of sovereignty in the South and East China Seas. In these situations,
leaders face a shared fundamental question about how to navigate a
dangerous situation without accidentally sliding over the line into a
larger-than-desired conflict (Snyder and Diesing 1977; Slantchev 2005).
Often leaders must make decisions with limited time and inadequate
information (Jervis 1976, p.~95). What guidance can we offer to help
maintain peace and stability? Can we predict how states will react to
specific events during crises? Can we recommend policies or actions that
are most likely to lead to success without being destabilizing? In
short, what empirical evidence can we bring to bear on the practical
questions of crisis management?\footnote{See Gannon (2022) for a recent
  review in the context of cross-domain deterrence and escalation.}

The international relations field offers two foundational perspectives
on crisis dynamics, classically described by Robert Jervis (1976) as the
deterrence model and the spiral model. The former blames escalation on
the failure to credibly communicate threats (Rovner 2020) or
misperceptions about capability or resolve (Morrow 2019). The latter
views escalation as a retaliation to provocation where a tit-for-tat
response fails to end the crisis (O'Neill 1991). These models offer
contradictory policy prescriptions: on one hand, policymakers should use
bold words and actions to discourage aggression; on the other, they
should mitigate security dilemmas through compromise and accommodation.
Historical studies typically provide more nuance, but the additional
context needed to make sense of specific crises just highlights the fact
that both models are underspecified. Neither model, moreover, is well
integrated with more recent thinking about the underlying bargaining
dynamics of military contests (Fearon 1995). Previous empirical research
designed to validate or reject competing theoretical explanations for
escalation has faced daunting challenges given that the basis for
comparison and/or the logical discrepancies between alternative theories
have not been resolved (Glaser 1992; Kydd 1997; Zagare and Kilgour 1998)
and the ability to empirically assess such efforts is complicated by the
structural indeterminacy of uncertainty (Gartzke 1999; Chen, et
al.~2022). These efforts have yet to establish a consensus pecking order
between canonical alternatives (c.f., Wright 1965; Carlson 1995;
Kinsella and Russett 2002).

Recognizing the enormous efforts of previous theoretical and empirical
scholarship, we adopt an abductive approach designed to explore the
correlates of escalation/de-escalation in a novel data source
constructed by the authors and others (Douglass, et al.~2022; Douglass,
et al.~forthcoming). Our approach capitalizes on a new dataset, the
International Crisis Behavior Events (ICBe) dataset, augmented by
machine-coded data ICBeLLM. To our knowledge, these are the first
international conflict data to code the full sequence of statements and
actions by actors necessary to capture escalation as a process. These
data allow us to track the actions and reactions of every relevant actor
in the 475 interstate crises available from the canonical ICB dataset;
who did what to whom, when, where, and how. Doing so allows us to
identify various factors most likely to lead to escalation or
de-escalation in crises. Using these triggers, we can develop a model
that estimates (predicts) both the path and duration of international
crises and that indicates likely precursors and their effects. We find
support for neither the deterrence nor spiral model, in their classic
forms, suggesting the need for a more nuanced theory that can
accommodate the historical contingency and complexity of crisis
processes.

This analysis is organized as follows. Section 2 briefly formalizes the
concept of an international crisis and the concept of escalation within
a crisis. It then draws prior beliefs from existing theories of
escalation, particularly deterrence and spiral models of conflict.
Section 3 introduces the data, turn-by-turn event data collected for 475
individual crises (Douglass, et al.~2023). This section also presents
our estimand---the unobserved latent quality of an action being
escalatory or de-escalatory at a specific point in a crisis---and our
measurement strategy, which involves observing the empirical increase or
decrease in the remaining duration of a crisis as a function of the
observed actions. Section 4 describes our modeling strategy and presents
empirical results. Section 5 provides a qualitative deep dive with a few
choice case studies (Tavory and Timmermans 2014; Heckman and Singer
2017). Section 6 offers concluding remarks and implications of our
findings.

\section{Crises, Escalation, and the State of the
Art}\label{crises-escalation-and-the-state-of-the-art}

The scientific state of the art on crisis escalation is too large and
amorphous of a topic to fully summarize, and so we draw attention to the
minimal necessary machinery for understanding the empirical argument
made here. Those components are the domain of the activity in question,
the definition of the outcome we seek to explain, theoretical priors
about the connection between different possible actions and that
outcome, and empirical attempts to measure the above.

\subsection{Crisis and Stability}\label{crisis-and-stability}

The status quo in international relations is almost always one of
stability, slow-changing bargains, and relative cooperation. War and
conflict are the exceptions (Vasquez and Valeriano 2010). Formally,
consider a set of autonomous political actors or players, \(p\in P\),
consisting of states, international organizations, and subnational
organizations. At each moment in time, \(t\), they engage in behaviors,
\(b\in B\), individually and towards each other which they each then
hold preferences over. Any player can transition to a period of crisis
when those behaviors swing toward ``disruptive interaction with a
heightened probability of military hostilities that destabilizes states'
relationships or challenges the structure of the international system''
(Brecher and Wilkenfeld 1982). The International Crisis Behavior project
has documented 496 such periods since 1918 (Brecher and Wilkenfeld
1997). Formally, we can think of this as a simple Markov model with two
discrete latent states, stability \(S\) and crisis \(C\),
\(\Theta_{p,t}\in[S,C]\). This latent state is in turn partially
observed in the current and historical behavior of the players.

The measurement strategy for the unobserved latent state is to use
observables like behaviors as proxies, \(b_{p^-1,t}=F(\Theta_{p,t})\).
When other players are behaving badly, we think a player is strictly
more likely to be in a state of crisis. However, the policy-relevant
question we want to answer concerns the causal impact of a behavior on
the continuation or resolution of a crisis,
\(\Omega_{p,t}=F(b_{p^-1,t-1},b_{p,t-1})\). This reveals an immediate
conceptual difficulty. Behaviors appear on both sides of the regression
equation, first on the left hand side as a proxy measurement for the
dependent variable of state of crisis in time \(t\) and then
individually on the right hand side as potential causes of crisis if
observed in time \(t-1\). This means the definition and measurement of a
discrete-time step within a crisis is doing the majority of our
conceptual work. This fact has created unending confusion between
measurement, explanation, prediction, and prescription in the study of
crises, and we tackle it directly below.

\subsection{Defining Crisis}\label{defining-crisis}

For now, imagine a stylized crisis with two discrete and clearly
distinguishable time steps, \(t\) and \(t-1\). In \(t\), we want to know
whether a player is in a state of stability or crisis. We observe the
behaviors of other actors in time \(t\), \(b_{p^-1,t}\). International
relations theory holds strong beliefs about which behaviors are more
crisis-like. For each behavior, \(b\), assign a weight
\(w=[w_1,w_2,...,w_n]\), representing the strength of the signal it
provides that a player is currently in a crisis state and the discrete
state is some function of the sum of each of these weighted observed
behaviors \(\Theta_{p,t}=F(\sum^n_{i=1}w_i*b_i)\).

We draw prior beliefs from international relations literature about the
sign and rank ordering of these weights across behaviors. Much of
20th-century security studies was preoccupied with conventional warfare
as the most extreme behavior, e.g.~the Correlates of War (COW) project.
In the shadow of possible nuclear war, scholars turned to measuring
events that might be a precursor to actual fighting, e.g.~the threat,
display, or use of military force short of war recorded by the
Militarized Interstate Disputes (MIDs) dataset (Maoz, et al.~2019).
Others recorded entire streams of events from news headlines (McClelland
1978; Goldstein 1992; Azar 1993; Sherman 1994) and defined their level
of departure from stability by hand, e.g.~Goldstein scores (Goldstein
1992), or through unsupervised dimensionality reduction, e.g.~Item
Response Scores (Schrodt 2007). The ICB project took an even wider view,
recording a ``trigger'' for each crisis which could take one of 9
different forms (verbal act, political act, economic act, external
change, other non-violent act, internal verbal or physical challenge to
regime or elite, non-violent military act, indirect violent act, or
violent act).

The ICBe ontology divided crisis events into two conceptual categories
across armed actors vs unarmed actors and, given that,
escalatory/uncooperative vs de-escalatory/cooperative behavior, shown in
Table 1 below. It further recorded details related to scale and
severity, like the number of troops involved in an act, number of
casualties, amount of territory exchanged, etc. The labels of
escalatory/de-escalatory were perhaps premature, and instead for our
purposes here we argue these are better thought of as indicators for and
against a state of crisis/stability within a single discrete time step.

\includegraphics{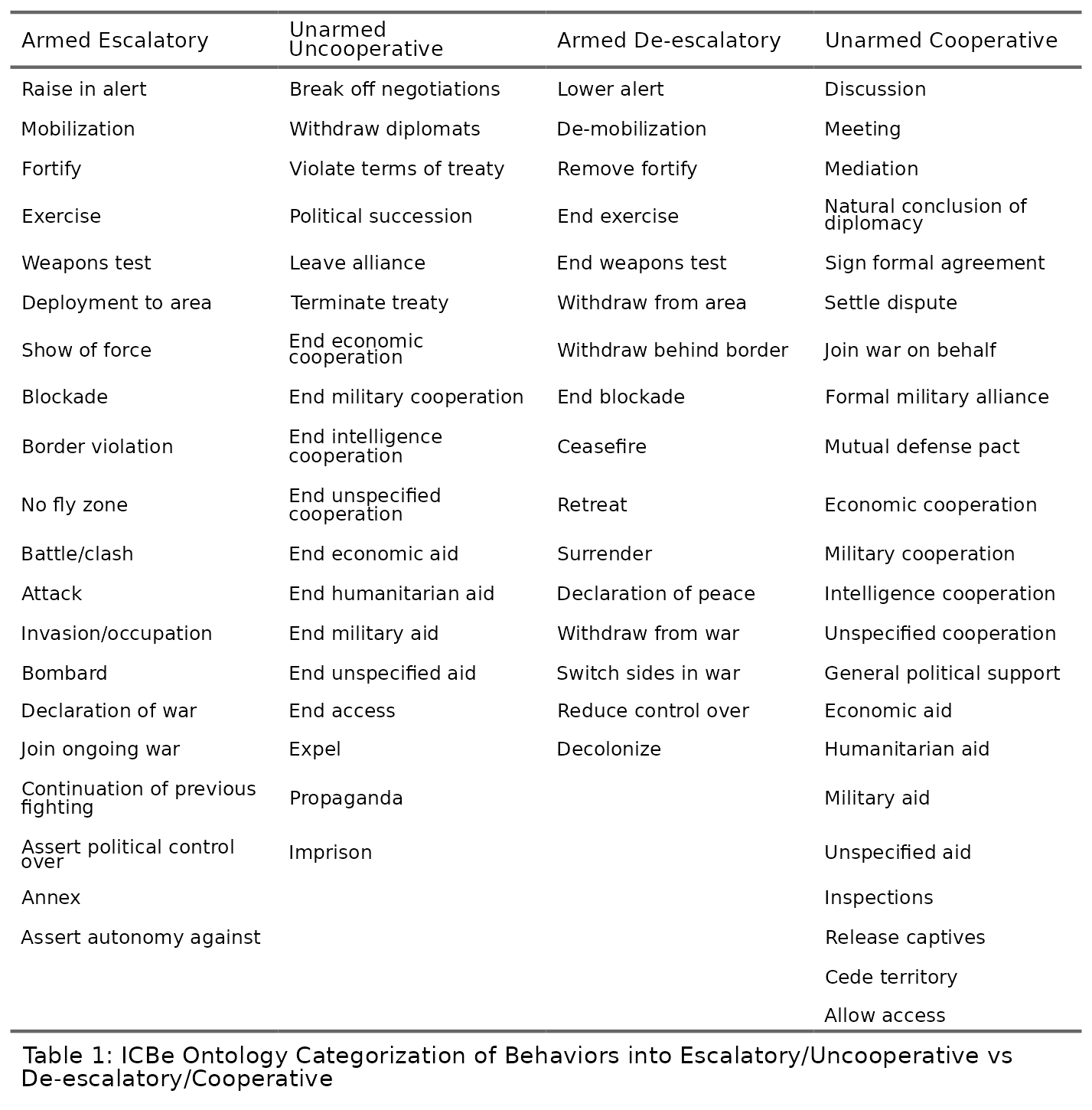}

\subsection{Defining
Escalation/De-escalation}\label{defining-escalationde-escalation}

We take exactly one time step back in our stylized crisis to \(t-1\) and
define the relationship between a given behavior and crisis stability.
For simplicity, we assume the player is already in a state of crisis at
\(t-1\), and so the only transition probabilities we are interested in
are from crisis to crisis, \(P(C\rightarrow C)\), and from crisis to
stability \(P(C\rightarrow S)\). Because the two outcomes are mutually
exclusive, we can further consider just a single transition, and so we
choose \(P(C\rightarrow C)\), the probability of crisis being followed
by more crisis. As before, we can think of all of the same behaviors,
\(b_{p^-1,t}\), simply observed one time step earlier, and we can assign
another weight vector \(\beta=1,2,..,n\), and specify some function that
maps the last period's behaviors to the next period's probability of
being in a state of crisis
\(\Theta_{p,t} = F(\sum^n_{i=1}\beta_i * b_{t-1,i})\).

Unlike before, the theoretical literature on crisis stability is far
more divided on what those weights should be (Brams and Kilgour 1987;
Powell 1989). As a gross simplification, one major vein of literature
views the weights of escalation to be highly correlated with the weights
indicating a crisis (Colaresi and Thompson 2002), while another views at
least some of the weights as flipped (Spaniel and Idrisoğlu 2023).

\subsection{Perspectives on
Escalation}\label{perspectives-on-escalation}

One might assume that doing more aggressive and destructive things would
be followed by even more aggression and destruction. Yet those same
aggressive actions might intimidate others into backing down. In one
case, aggression begets aggression, and in the other, aggression
encourages submission. Security dilemma models following from
pessimistic assumptions tend to focus on the first case, and relatively
more optimistic deterrence models on the second. Scholars have explored
subtle situations in which even two well-meaning players might find
themselves caught in a conflict that neither of them intended. The
so-called spiral model considers tough talk and a forward military
posture to be inflammatory, exacerbating security dilemmas and
heightening the risk of war (Herz 1950; Smoke 1977; Jervis 1978; Glaser
1997). Threats alert an adversary, weakening defense and triggering
anticipatory reactions. A nation targeted by an adversary with threats
of retaliation should be less likely to back down if there are
reputational costs (domestic, international) or psychological burdens
for failing to fight (Jervis, Lebow, and Stein 1989; Lebow and Stein
1989). Indeed, such threats may trigger action where none exists. In the
model above, the weights of escalation and crisis indicators are highly
correlated.

Another vein of literature argues that at least some of the weights
should be flipped. If weakness or uncertainty invites conflict, then
some of the most crisis-like behaviors may in fact be needed to
transition back to stability. Deterrence theory fundamentally sees
conflict onset, and by extension escalation, as a cost-benefit analysis
(Freeman 2004). Nations that find war prohibitive, or that are unwilling
to endure the risks of escalation, are unlikely to fight (Morgan 1977).
Deterrence is most effective when threats are clear, plausible, and
consequential---when words credibly signal contingent actions (Powell
1990). Deterrence theory thus advocates the use of threats to impose
costs or risks on a potential adversary in the hopes that the prospect
of risk or harm will derail aggression and discourage escalation (Brodie
1959; Wohlsletter 1959; Snyder 1961; Schelling 1966; Waltz 1990;
Danilovic 2002). The empirical challenge in identifying deterrence
success given strategic self-selection into disputes (Fearon 2002) has
motivated others to disaggregate this idea further, introducing
distinctions between, for instance, immediate deterrence, where a state
or other actor is seeking to prevent an actual attack from an adversary,
and general deterrence, where the effort to discourage an adversary from
even considering force as an option (Huth 1988). Many other such nuances
have been employed, highlighting the underappreciated role of historical
context in determining the escalatory dynamics of any given crisis
(Kreps and Schneider 2019; Cunningham 2020; Lin-Greenberg 2023).

Despite the better part of a century of speculation and debate, experts
have been unable to establish a consensus theoretical understanding of
escalation, its causes, or likely triggers (Kahn 1965; Kydd 1997; Smith
1999; Sechser and Fuhrmann 2013). Fundamental disagreement persists
about what words or actions are escalatory and which are more likely to
reduce tensions between actors. Other research questions whether crisis
actions such as threats and mobilizations are even a primary driver of
crisis dynamics or if circumstantial factors, such as power and
distance, are sufficient to account for actions and outcomes (see Stam
1996, Schultz 1998, Smith 1998, Gelpi and Griesdorf 2001, Kinsella \&
Russett 2002, and Gartzke \& Hewitt 2010 on interests, Huth 1988 and
Mearsheimer 1983 on capabilities, Banks 1990, Slantchev 2010, and Fey \&
Ramsay 2011 on information, Gartzke 1999 and Chen et al.~2022 on
uncertainty, and Huth, Gelpi, \& Bennett 1993 on nuclear weapons).
Scholars have begun to make sense of various conflict processes,
creating the opportunity to develop tools capable of predicting and
assisting in the evaluation of policy options (e.g., Hegre et al.~2017).
However, contemporary theories fall short of providing clear, consensus
perspectives on escalation management (Altfeld 1983, Diehl 1985,
Nalebuff 1986, Moul 1988, Fearon 1994, Carlson 1995, Bueno de Mesquita,
et al.~1997, Reed 2000, Huth and Allee 2002). Ambiguity is rooted in the
conflicting spiral and deterrence frameworks that have developed to
explain crisis dynamics.

Both the classical spiral and deterrence frameworks are rooted in
earlier intuitions about the nature of war. More recently scholars have
come to appreciate the centrality of bargaining in conflict processes
(Fearon 1995). In this view, actors are strictly better off coming to a
mutually advantageous deal rather than engaging in costly conflict,
which destroys the bargaining surplus, but they also have incentives to
misrepresent their capabilities or intentions to get a better deal
without fighting. Yet, this means that there is an important endogenous
role for bargaining moves in a crisis that can either resolve or
exacerbate information problems (Gartzke 1999; Coletta and Gartzke
2003).

Process matters, as historians have long understood, but the spiral and
deterrence models do not explicitly incorporate contingency. Process
seems to be more relevant in the spiral model, which implies a long
history of tit-for-tat interactions building up to an explosive crisis,
yet even here, each event is assumed to be positively correlated with a
greater likelihood of war in the future, so the temporal sequence is
less relevant than it first appears. The deterrence model, likewise,
cannot explain why the same type of move, say the mobilization of troops
on a border or deployment of warships to a coast, is followed by
stability in one crisis yet instability in subsequent crises. The string
of crises leading up to World War I is only the most obvious example. To
use the terms of the illustrative model above, there is some other
vector of \(X\) contextual conditions that determines whether escalatory
weights are correlated one way or the other. Classical theories are
underdetermined because they've just taken a coarse average over all of
those conditions.

\subsection{The Measurement Problem in Escalation
Research}\label{the-measurement-problem-in-escalation-research}

At least part of the difficulty with making sense of escalation
empirically lies in the limitations imposed by existing data.
Conventional datasets in international security either code outcomes or
events, which means that the conflict processes modeled above are
aggregated into discrete data points. However, escalation is a process,
a series of relationships that both tie events together across time and
that are sequentially meaningful or significant. Immediate deterrence
implies a general deterrence failure (perhaps). A spiral requires
triggers and effects, each related to the other. If researchers lack
information about what might be called the ``begats'' of international
relations---what factors preceded and precipitated outcomes of
interest---then it will be difficult to decide whether deterrence or a
spiral even occurred, let alone how they work and why.

There are generally two types of datasets that code conflict behaviors
in international relations, conflict and events data. Conflict data
provides information about outcomes. Events of interest such as wars,
disputes, or other behaviors are recorded, with dates and other relevant
and useful information (actors, initiator/target, intensity, location,
etc.) included to aid researchers in conducting analysis. A prominent
example of conflict data is the Correlates of War dataset, especially
the Militarized Interstate Disputes (MIDs) dataset (Maoz, et al.~2019).
While far from perfect, these data have been utilized in hundreds of
studies over decades. They are considered fairly reliable. However, they
do not specifically delineate why outcomes of interest occurred. Indeed,
this is generally the subject of conjecture (hypothesizing) and testing
using these data. It is especially difficult to discern whether given
outcomes are the product of other acts or behaviors, such as is assumed
to occur with escalation and other conflict processes. A standard
practice is actually to use statistical techniques intended to eliminate
any correlation between events in these data, due to concerns about
common priors (dependence) (Beck, et al.~1998; Carter and Signorino
2010). Because of the structure and coding of conflict data, it is thus
extremely difficult to effectively assess claims about processes, such
as escalation, where values of given variables in the dataset are
assessed to discern whether they lead to other values.

A second, less common though important, and innovative form of conflict
data collection involves events datasets. Events data have a long
history in international relations (McClelland 1978; Goldstein 1992;
Azar 1993; Sherman 1994), which has ebbed and flowed over time as new
techniques have been introduced and as theoretical and substantive
interests evolved. In essence, the promise of events data is the ability
to capture precisely what conflict data is missing, the ``begets'' in
world affairs (Merritt, et al.~1993). By collecting a large and,
increasingly with the rise in computing power, vast amounts of events
the expectation was that the bonds between events would become apparent
also. Projects such as CAMEO (Schrodt 2012), ACLED (Raleigh 2010), and
others have managed to hoover up enormous quantities of data. The
problem has been what to do with it, and how to make it make sense of
the world.

As it turns out, there may not be a direct connection between the amount
of information (number of events) collected and the ability to capture
the ``begats'' that are intrinsic to making sense of processes. We just
do not know in a large corpus of conflict events what prior events do to
make subsequent events more or less likely. Lacking clear theory or
correlates, the tendency has been to ``bin'' events, creating counts by
type or intensity by time period that begin to look much more like
conflict data than was originally the intent of events researchers. We
propose a different, hybrid approach seeking to sort out cause and
effect in studying escalation, at least as an interim measure.

\section{Data and Measurement
Strategy}\label{data-and-measurement-strategy}

We employ narratives of 475 crises recorded by the ICB project version
12. To measure the actions that take place within those crises we employ
the International Crisis Behavior Events dataset (ICBe) (Douglass et
al.~2022). Human codings are provided by ICBe v1.1, which we subset to
just expert-trained coders (discarding additional novice student
coders). Machine codings are provided by ICBeLLM (Douglass et al.~2023),
which uses a large language model with engineered prompts to approximate
the human application of the ICBe ontology to crisis texts. To maximize
precision, we keep only details that were reported by at least two
experts or by one expert and the LLM. The raw unit of analysis is the
crisis-sentence-actor set (n=15,487). An actor-set is a unique
combination of values for \{think\_actor\_a, say\_actor\_a,
say\_actor\_b, do\_actor\_a, do\_actor\_b\}. Each sentence can then be
mapped to multiple events as long as those events involve the same
actors (e.g.~an action by actor A against actor B can be coded alongside
a different action by actor B against actor A, and a speech act can be
coded in addition to a separate action, etc.). The inclusion of ICBeLLM
increases the number of observations from 9,358 expert human coder-only
events to 15,487 events total, a 65\% increase in the sample size
without compromising quality and without hiring and training additional
human coders.

\subsection{Unit of Analysis}\label{unit-of-analysis}

Our unit of analysis is the crisis-time step. We have 475 crises. Our
stylized example crisis had only two time steps, whereas we estimate
between 2 and 68 time-steps (median 13) for our corpus of narratives.
There are 7,126 crisis-time steps in total. A time-step is defined as a
sentence in a narrative that generated any ICBe/ICBeLLM events. We
aggregate all behaviors observed across events within that time-step,
e.g.~a sentence describing a threat by one actor and an attack by
another are all assigned to that single time step.

\subsection{Measuring Crisis -- ICB Universe of
Crises}\label{measuring-crisis-icb-universe-of-crises}

We rely on the human judgments of the ICB project to define the start
and stop of a crisis. We imagine the qualitative coding process as
dragging a convolutional filter across a sequence of behaviors and then
firing once the perceived departure from the norm reaches sufficient
strength. When the ICB narrative stopped describing events for that
crisis we consider that a transition from crisis to stability.

\subsection{Measuring Escalation/De-escalation -- Time Until Crisis
End}\label{measuring-escalationde-escalation-time-until-crisis-end}

Our outcome of interest is the number of time steps remaining until a
given crisis ends, \(Y_{c,t}\in [0,...,max(t_c)]\). Our stylized example
crisis consisted of only two time steps and a single transition. Our
full crises consist of multiple steps, any one of which could have
transitioned to stability. The number of steps remaining provides more
information than a single yes/no transition, and it has a nice intuitive
interpretation. `Time until crises end' is a directly interpretable
proxy measure of escalation/de-escalation. A behavior that greatly
increases the expected length of the crisis is escalatory and one that
shortens the length is de-escalatory. This definition does not rely on
pre-decided judgments of behaviors, and it allows for theories that
argue even more aggressive/violent behaviors might actually shorten or
deter a crisis altogether.

\section{Modeling Strategy}\label{modeling-strategy}

Our ultimate goal is to be able to give policy decision-makers
evidence-based recommendations about what kinds of actions they can take
to de-escalate a crisis and restore a stable international environment.
Unfortunately, we lack the leverage to establish the necessary
counterfactuals to make such strong causal claims. However, with a large
and increasingly detailed historical repository of international events,
we can achieve a related goal of giving policymakers the analytical
ability to see how an unfolding crisis compares with other crises,
especially those that are objectively ``most similar'' within a given
period of history. For any currently unfolding crisis, we should be able
to say that given events up until a certain point in time \(t\) are the
most valid historical analogies with similar histories. Although history
never repeats itself exactly, knowing the distribution of outcomes among
an event's nearest rhymes or whether an ongoing crisis has no near
historical analogs is itself useful for policymakers.

\subsection{When Behaviors Appear in
Crises}\label{when-behaviors-appear-in-crises}

To illustrate the idea, we begin first with a simple descriptive
exercise arranging the types of behaviors we observe empirically in
crises by when (i.e.~the time point) that they tend to occur within a
crisis. Our unit of analysis is the crisis-sentence-behavior. We
calculate each behavior's relative position within a crisis by counting
the number of behaviors occurring before it and dividing this number by
the total number of events, \(Y∈[0,1]\). The mean location within
narratives across the entire corpus is shown in Figure 1 below.

\includegraphics{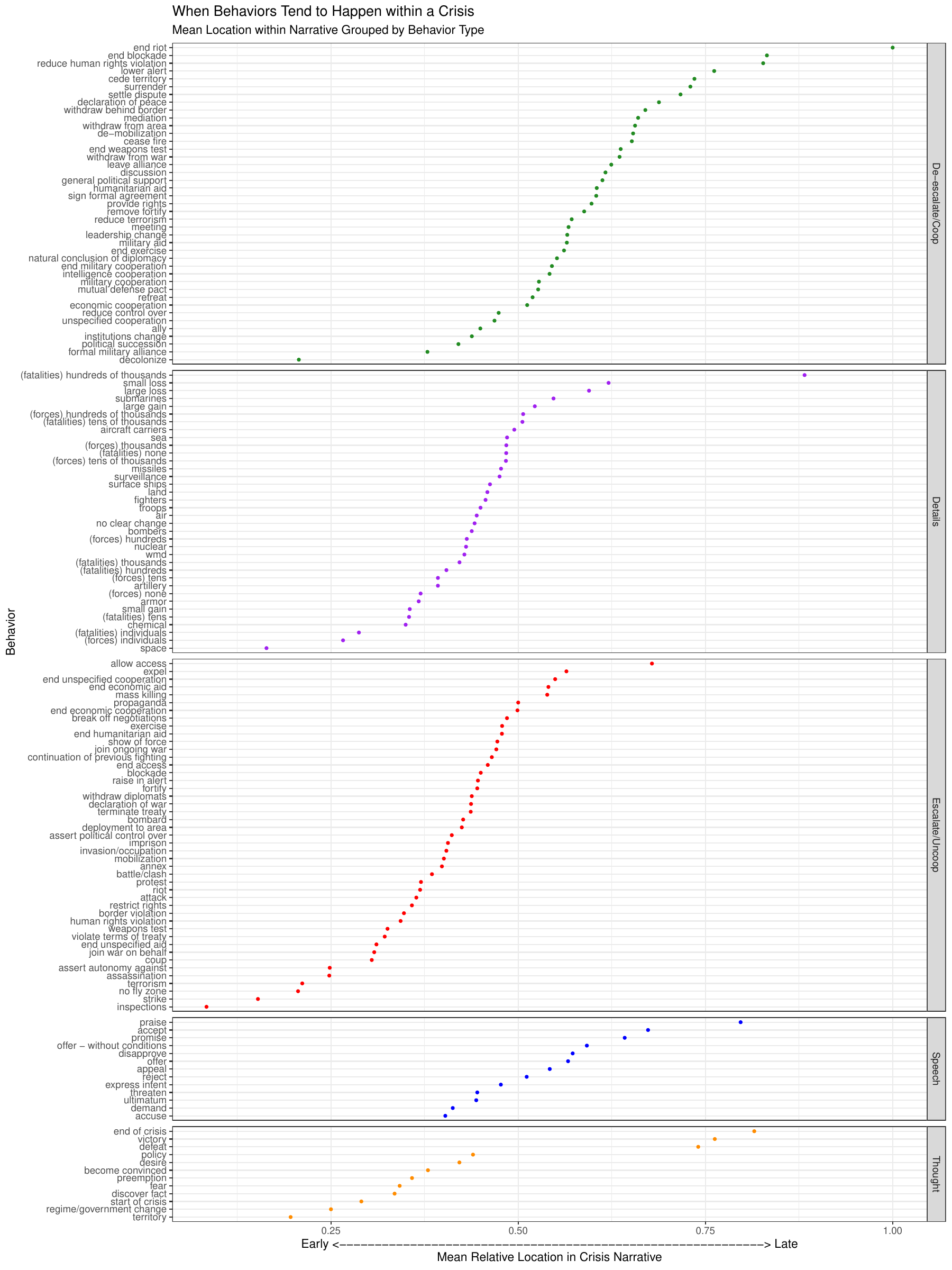}

We find that behaviors align in time in a way that largely mirrors our
prior beliefs about what constitutes escalatory/non-cooperative versus
de-escalatory/cooperative behaviors. The ICBe ontology explicitly binned
actions into escalator/non-cooperative (red)---which we find here appear
earlier in crises---and de-escalatory/cooperative (green)---which we
find appear later in crises. The two groups are so strongly correlated
with time within crises that they are nearly linearly separable. Thought
behaviors were not pre-divided into types, but they similarly align over
time with motivations for conflict appearing early (start of crisis,
territory aims, regime change aims, preemptive aims, discovering or
becoming convinced of a fact, etc.) and thoughts ending late
(perceptions of defeat, perception of end of a crisis). Likewise, speech
acts that appear early on in crises include accusations, demands,
ultimatums, and threats, while speech acts that tend to be associated
with the ending of crises include promises, offers, and acceptances. The
details of actions tend to appear earlier in crises, e.g.~how many
troops, while the consequences, e.g.~changes in territory, tend to come
toward the end.

\subsection{Predicting Time Until Crisis
Resolution}\label{predicting-time-until-crisis-resolution}

We can now move to analyzing variation within crises. Our unit of
analysis shifts to the crisis-time step, where each sentence of the
crisis narrative with coded ICBe events represents a single step. This
measure (number of time steps remaining until the termination of a given
crisis) roughly represents our earlier discussion of crisis resolution
versus extension. When a crisis begins at \(t=0\), participants do not
know whether it is going to last a long time or end quickly. Once we've
seen a few behaviors, we want to update our beliefs and ask whether it
is more likely that participants are getting closer to, or conversely
further away from a resolution of the crisis. Importantly, we want to
understand which behaviors signal the likely impending termination of a
crisis.

We want to predict the time it will take to resolve a crisis as a
function of the history of behaviors until time \(t\),
\(Y_{c,t}=F(X_t)\). We encode the history of crisis behaviors as
features of the crisis in the following way. For each of the 140
possible behaviors/details, we calculate the number of time steps since
it was last observed (0 in cases where it appeared in the current time
step, 1 for one time step ago, etc.). For behaviors/details not observed
at all, we set them at an extreme value of 100, allowing our tree-based
estimator to intelligently bin events that never appeared with those
that appeared a very long time ago, or alternatively to split them off
entirely as a different category of behaviors altogether. We further add
a single control variable of the crisis number coarsened into five
blocks of \textasciitilde100, to account for the ICB crisis project
trending toward writing longer narratives for more recent years.

Our cross-validation strategy takes into account the nested-panel
structure of our data. We divide the \(C=475\) crises into 10 test
folds. For each test fold (\(c=48\)), we fit a model dividing the
remaining crises, \(c=427\), into a training set of crises, \(c=405\),
and a validation data set, \(c=23\). The same crisis will therefore
never appear in both training and test, or training and validation, etc.
Because we fit 10 different models predicting out-of-sample for each of
the 10 test folds, we are able to present results below that cover the
entire dataset and are based only on out of sample predictions.

We employ a gradient-boosted trees estimator, specifically LightGBM (Ke,
et al.~2017). We choose this estimator because of its speed,
flexibility, and predictive performance. It is nonparametric, allowing
us to easily distinguish differential effects of a behavior happening
recently from long ago, and to automatically search for interactions
among behavior types. We optimize an L1 regression loss, mean absolute
error, rather than mean squared error, which is less sensitive to
outlier very long crisis narratives, and empirically had better
out-of-sample performance.

Our estimand is the degree to which previous behaviors lengthen or
shorten a crisis, the degree to which a behavior is escalatory, the
weights \(\beta\) introduced earlier. We approximate that estimand using
the out of sample feature contributions of each behavior - time since
observed. Our measure of feature contributions is SHAP values which is a
local approximate decomposition of each feature's contribution to the
final prediction produced by the full black box model.

The results are shown in Figure 2 below. The outcome is predicted time
until the end of the crisis (defined as no more events coded from the
narrative). Along the Y-axis each row lists the behaviors with the
greatest contribution to the model's predictions. Along the X-axis, each
column lists the number of time steps in the past that behavior was last
observed, with 0 being concurrently to a given time \(t\). The cells are
shaded with the observed SHAP value for that behavior last observed that
many time steps ago, with bright red reflecting a strong increase in the
expected number of steps until the end of the crisis and deep blue
representing a strong expected decrease in the number of steps until the
end of a crisis. Behaviors are arranged in descending order, from the
largest expected increase in time until crisis end to the largest
expected decrease (sum each row's total SHAP contribution, dividing each
cell's value by the time since it was observed to place greater weight
on more recent acts).

\includegraphics{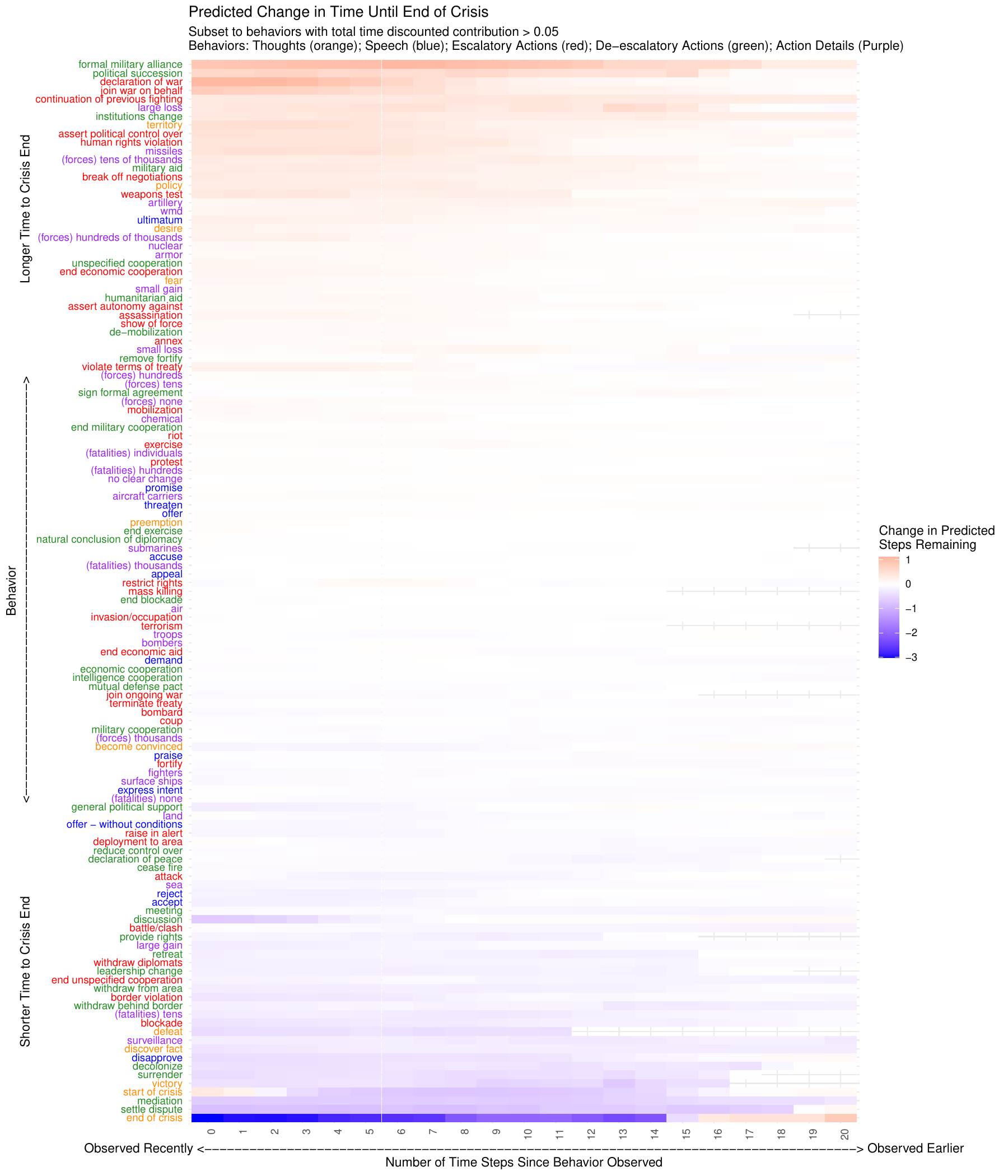}

We start by describing the easiest-to-understand behavior, ``end of
crisis,'' which is something that was explicitly mentioned in each
crisis narrative and coded by ICBe as a thought behavior. Once the
narrative mentions that one of the actors has decided that the crisis
has ended, the model expects the whole narrative to wrap up shortly and
stop producing coded events. The effect also grows stronger in time; the
longer in the past that an end to the crisis is mentioned, the more
likely it is that the narrative really will be wrapping up soon, until
at some point the sign flips and it becomes more likely that the crisis
ended for only a subset of actors and the rest continued on their own
path. On the opposite end of the spectrum, forming a military alliance
is most indicative that the crisis is going to extend longer and as
before the strength is greatest the more recent the behavior was
observed.

Zooming out, several general trends are immediately apparent. First,
there was a strong signal for only some behaviors with many having
little or no contribution to the model's predictions (we show 123 of 140
here). This is a function of the actual data-generating process (some
behaviors are orthogonal or redundant to crisis duration), to
measurement (some behaviors are measured or reported more noisily than
others), and to the sample size (a larger corpus would likely tease out
the role of more behaviors). Second, the effect of a behavior generally
decays in time, with more recent events being more informative than
those that are more distant in time. Third, there is variation in the
degree of decay across behaviors. Figure 2 depicts a triangle-shaped
surface, with behaviors toward the middle still being informative but
only if they have occurred relatively recently (the shape is an
intentional artifact of the ordering).

Moving to a substantive interpretation, the types of behaviors that
extend versus those that reduce the length of a crisis lend support to
some theories of international relations more than others. We find the
biggest winners to be bargaining models of war (Fearon 1995; Wagner
2000, 2007) that see large violent shocks as stability-inducing because
they provide information to both sides about who is likely to win should
the contest continue (Shirkey 2016, Weisinger 2013). Behaviors that
indicate a crisis is coming to an end include a perception of victory or
defeat, a surrender or withdrawal from an area or a retreat, a large
gain in territory, and to a lesser degree even battles/clashes and
attacks. Meanwhile behaviors that indicate a lengthening of the crisis
are large changes in the international system that might increase
uncertainty, like the formation of a new military alliance (Benson 2012;
Benson and Smith 2022), a political succession, a declaration of war, a
third party entering a way (Shirkey 2012), or a change in institutions
change.

In terms of diplomacy, aggressive speech that might signal resolve like
making an ultimatum or breaking off negotiations, are more likely
followed longer crisis periods. Meanwhile engaging in mediation, making
an offer without conditions, even expressing disapproval, rejection, or
acceptance, all indicate the crisis is shifting to speech and away from
actions and likely to end sooner.\footnote{These could be associated
  with the end of a crisis because the structure of the narratives
  describes mediation efforts at the conclusion of the crisis summary.
  Later analyses could identify the time period when mediation efforts
  occurred relative to the observed end of the crisis to see if
  different patterns emerge.} Mentions of grievances and specific topics
are also indicators of more time until resolution, such as threats or
accusations, fears or desires, territorial or policy aims, becoming
convinced of a policy fact, human rights violations, etc.

We stress that this is a purely observational research design that
cannot distinguish causes of crisis resolution from coincident
indicators of a crisis winding down. For example, mediation may not
actually cause a crisis to end. Rather, mediation may be associated with
the end of a conflict process that's already begun to terminate for
other reasons. That said, we do find unexpected correlates in behaviors
we considered escalatory/un-cooperative being associated with a crisis
ending sooner and others that we have considered
de-escalatory/cooperative being associated with a crisis persisting for
longer periods of time. In particular, the ones that flipped appear to
be consistent with a bargaining model view of international conflict,
with cooperative acts that signal major realignments---and thus create
additional uncertainty---extending crises, while escalator/uncooperative
acts that kill and destroy but provide information being more often
associated with bringing a crisis to a close.

From the perspective of the debate between deterrence and the spiral
model, these results are much less supportive. Preparing to fight a war
and actually declaring it are closely related, and so mobilizing forces,
events involving tens of thousands of troops, conducting a weapons test,
and other similar events that potentially transmit information about
resolve (signaling) all tend to indicate that the crisis is just warming
up. Likewise, intensive fighting and to a lesser degree just fighting in
general tend to indicate a crisis is coming to an end. As a mental
model, this corpus of crises can be thought of as long lead-ups in
grievances and disputes ending either in diplomacy or exploding in the
use of decisive violence. To the degree that decisive violence is
self-perpetuating, it is more likely to be evidenced in the creation of
new crises between the same actors over years as they regroup and find a
new approach/vector to attack one another.

\section{Case Examples}\label{case-examples}

In this section, we detail three case study examples that help to
illustrate how behaviors impact expectations over the course of a
crisis. Additional details of each case are depicted in Figure 3. The
Y-axis in Figure 3 reports the predicted number of steps estimated to
remain in the crisis. The X-axis details the step T in the crisis (note
that the length of each crisis differs). We annotate each point with the
behavior that most strongly contributed to a given prediction (colored
by the strength and direction of that particular contribution).

\includegraphics{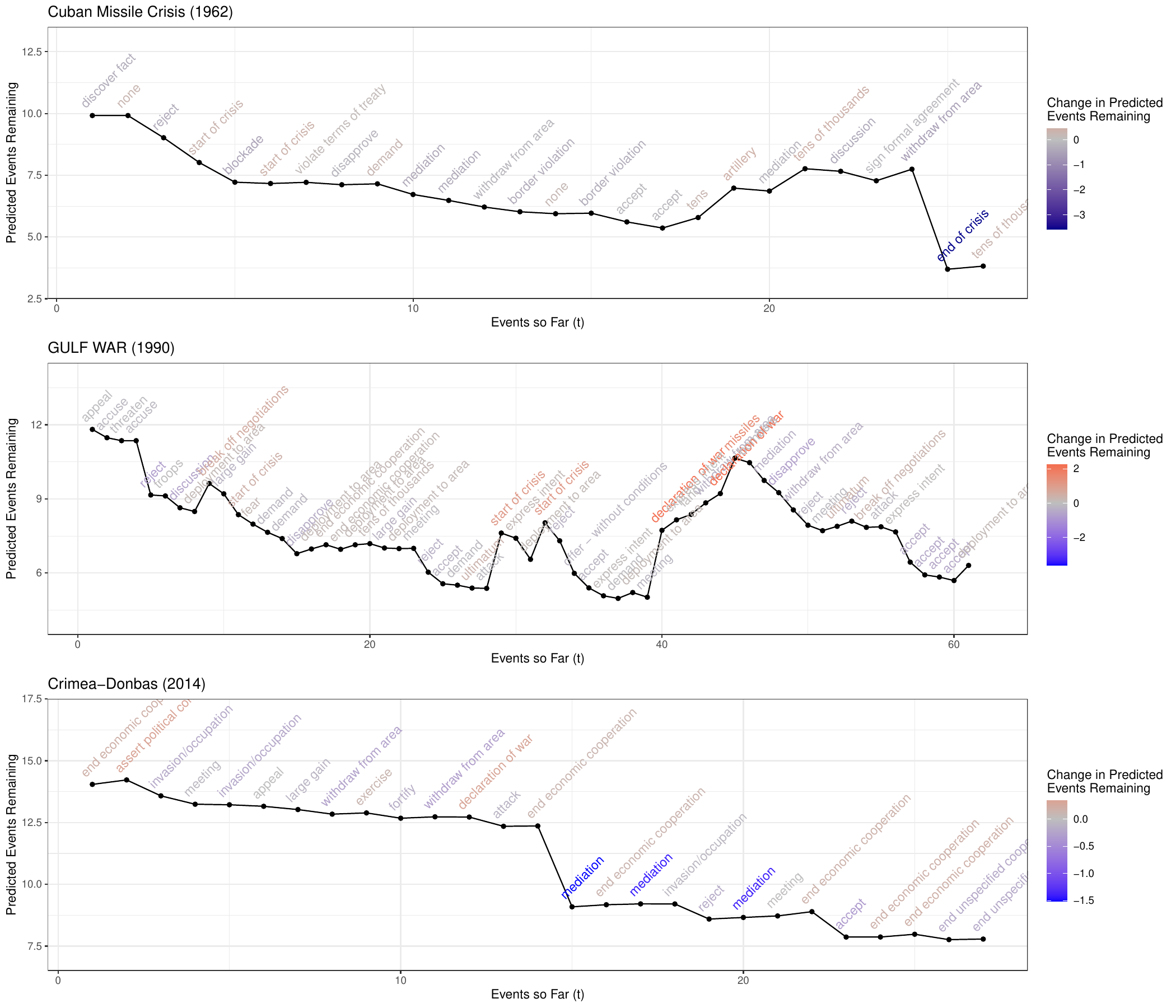}

For example, at the first time-step of the Cuban Missile Crisis, the
model estimates that there are likely to be 10 more steps, which is
driven primarily by the global average and partly by the only
information seen so far in the behaviors of just that first step. The
behavior in that first step that most contributed to the prediction was
that the United States discovered a fact, leading the model to increase
the expected length of the crisis. The next step contained little
information keeping the number of expected steps at 10 despite advancing
1. In the third step, the model lowers its expected length because of a
rejection combined with the mobilization last observed only 2 steps
prior. In this way, the model proceeds downward and to the right,
updating its expectation given the full history observed thus far. A
series of back-to-back promising signs, colored in blue, reduce the
predicted length of the crisis (e.g.~withdraw from an area, mediation,
mediation, withdraw), until the shootdown of the U-2 spy plane which
sparks an increase in the expected length of the conflict. It does not
begin to decline again until the signing a formal agreement, a withdrawl
from an area, and then an explicit end of crisis craters the expectation
to only 4 steps remaining. At some steps, the behaviors of the top
feature provide no clear signal, with small SHAP values (gray color).

The three example cases vary in their overall temporal signature. The
Cuban Missile Crisis is U-shaped with a high level of intensity early on
with a reversion right toward the end. The Gulf War has three distinct
phases, starting first with the invasion of Kuwait, which is followed by
furious negotiating and brinkmanship, which the model predicts will end
the crisis. In the second phase, the coalition makes an ultimatum to
invade and there is another round of negotiation. A final third phase
signals the start of the war. Finally, the Crimea-Donbas crisis began
with intensity but then stair stepped down once mediation between the
two sides began.

To see how representative these temporal trends are we next cluster the
predicted number of remaining steps at each time point for every crises,
shown in Figure 4 below. We find several general patterns similar to the
case studies above, across 6 clusters: crises that resolve almost
immediately, crises that begin to resolve but then revert to a high
unsatisfying level of hostility, crises that stair step down in
escalation, and finally crises that stair step upward over time. We
interpret this as evidence of intermittent rounds of fighting and
bargaining within crises. This demonstrates the importance of scale in
measurement. Some of these crises have episodes and breaks within them.
Earlier we noted that beyond a certain scale of conflict, events are
likely broken down across separate crises over years. How we interpret
the consequences of a particular behavior turns on our beliefs about the
importance of the short-, medium-, and long-term consequences of a
statement, action, or outcome. Future work will need to pay close
attention to the endogenous nature of how combatants define the start
and end of a crisis, and how social scientists in turn choose to group
events and periods into large constructs when conducting their analyses.

\includegraphics{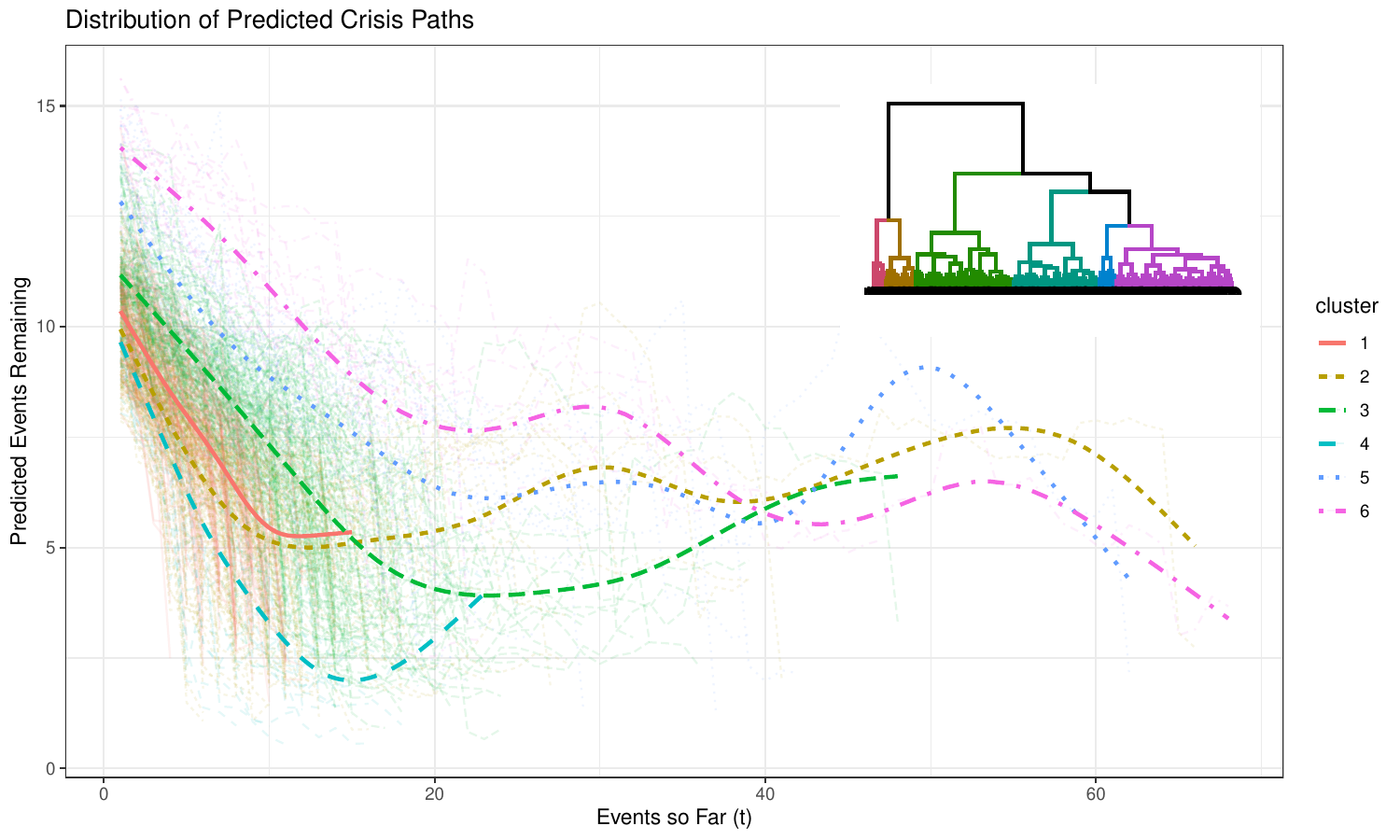}

\section{Discussion}\label{discussion}

The research design here is purely observational and will require
additional strong priors to make causal determinations from otherwise
correlational findings. That said, we provide two different views on
empirical regularities in crisis events that allow us to begin the
process of a more intensive, and detailed, disentangling of causal
claims. The first view lays out a large number of behaviors (140) on an
intensity scale from most to least escalatory as represented by whether
they tend to appear early in a blooming crisis (more escalatory) or
later when a crisis is winding down to its conclusion (less escalatory).
That proxy view very closely approximated our prior beliefs used in
building the original ICBe ontology and distinguishing
escalatory/uncooperative events from de-escalatory/cooperative events.
The second view we provide complicates the picture more and asks what
kinds of events are most indicative that a crisis is likely to end soon.
We find suggestive evidence that there may be certain types of events
that can shock a crisis into a quick conclusion - such as military
contests that result in territorial shifts, surrenders, or major
realignments through coups or a third-party joining in a war. Likewise,
we found suggestive evidence that certain kinds of behaviors are
associated with causing or extending a crisis, like changes in
international millitary alignment or shifts in internal institutions.

Much remains to be done to untangle relationships and to clarify cause
and effect. The relationships we identify are correlations, not
causations. But this may be sufficient if the goal is simply to
delineate factors that are associated with de-escalation, rather than
making specific causal claims. Much of our findings suggest that the
classical dialectic between deterrence and the spiral model is at best
broadly descriptive and has little predictive value. Deterrence may be
influencing behavior earlier on before a crisis manifests (i.e., general
deterrence). Consistent with other findings (Beardsley, et al.~2024), we
find little evidence in our analysis to support claims that immediate
deterrence works as predicted in theory; actors are not de-escalating
because an adversary escalated. At least, it is far from clear that this
dynamic is in any way modal. Instead, it appears more usual for states
to mirror image one another, escalating when the other does so, and vice
versa.

\section{References}\label{references}

Altfeld, Michael F. 1983. ``Arms Races -- and Escalation?: A Comment on
Wallace.'' \emph{International Studies Quarterly} 27(2), 225-231.

Azar, Edward E. ``Conflict and Peace Data Bank (COPDAB), 1948---1978.''
\emph{Inter-university Consortium for Political and Social Research
(ICPSR)} Codebook \#7767. University of Michigan.

Banks, Jeffrey S. 1990. ``Equilibrium Behavior in Crisis Bargaining
Games.'' \emph{American Journal of Political Science}, 599-614.

Beardsley, Kyle, Jonathan Wilkenfeld, Phuong Pham, Corinne DeFrancisci,
Diana Partridge, and Megan Rutter. forthcoming. ``Deterrence in the
Early Stages of Crisis'' In Kyle Beardsley, Patrick James, and Jonathan
Wilkenfeld (eds.), \emph{A Century of International Crises, 1918-2019}.
Ann Arbor: University of Michigan Press.

Beck, Neal, Jonathan Katz and Richard Tucker. 1998. ``Taking Time
Seriously:~ Time-Series---Cross-Section Analysis with a Binary Dependent
Variable.'' \emph{American Journal of Political Science}. 42(4):
1260---1288.~

Benson, Brett and Bradley C. Smith. 2022.~ ``Commitment Problems in
Alliance Formation.'' \emph{American Journal of Political Science}
67(4): 1012---1025.

Benson, Brett. 2012. \emph{Constructing International Security:
Alliances, Deterrence and the Moral Hazard}. Cambridge, UK: Cambridge
University Press.

Bismarck, Otto von. 2013{[}1899{]}. \emph{Bismarck: The Man and the
Statesman}, vols 1\&2. New York: Cosimo Classics.

Bosch, Richard. 2017.~ ``Conflict Escalation.'' \emph{The Oxford
Encyclopedia of International Studies}. pp.~1-26. Oxford, UK: Oxford
University Press.

Brams, Steven J., and Kilgour, D. Marc. 1987. ``Threat Escalation and
Crisis Stability: A Game-theoretic Analysis.'' \emph{American Political
Science Review} 81:833--50.

Brecher, Michael, and Jonathan Wilkenfeld. 1997. \emph{A Study in
Crisis}. Ann Arbor, MI: University of Michigan.~

Brecher, Michael, and Jonathan Wilkenfeld. 1982. ``Crises in World
Politics.'' \emph{World Politics} 34(3):380--417.

Brodie, Bernard. 1959. ``The Anatomy of Deterrence.'' \emph{World
Politics} 11(2): 173---191.

Bueno de Mesquita, Bruce, Morrow, James D., Zorick, Ethan. 1997.
``Capabilities, Perception and Escalation.'' \emph{American Political
Science Review} 91(1), 15-27.

Carlson, Lisa J. 1995. ``A Theory of Escalation and International
Conflict.'' \emph{Journal of Conflict Resolution} 39(3), 511-534.

Carter, David B. and Curtis S. Signorino. 2010. ``Back to the Future:
Modeling Time Dependence in Binary Data.'' \emph{Political Analysis}
18(3): 271---292.

Chen, Chong, Jordan Roberts, Shikshya Adhikari, Victor Asal, Kyle
Beardsley, Edward Gonzalez, Nakissa Jahanbani, Patrick James, Steven E
Lobell, Norrin M Ripsman, Scott Silverstone, and Anne van Wijk. 2022.
``Tipping Points: Challenges in Analyzing International Crisis
Escalation.'' \emph{International Studies Review} 24(3).

Colaresi, Michael P., and William R. Thompson. 2002. ``Hot Spots or Hot
Hands? Serial Crisis Behavior, Escalating Risks, and Rivalry.''
\emph{The Journal of Politics} 64(4): 1175--98.

Coletta, Damon, and Erik Gartzke. 2003. ``Testing War In the Error
Term.'' \emph{International Organization} 57(2): 445--48. doi:
10.1017/S0020818303572083.

Cunningham, Fiona S. 2020. ``The Maritime Rung on the Escalation Ladder:
Naval Blockades in a US-China Conflict.'' \emph{Security Studies} 29(4):
730--68.

Danilovic, Vesna. 2002. \emph{When the Stakes are High: Deterrence and
Conflict among Major Powers}. Ann Arbor, MI: University of Michigan
Press.

Diehl, Paul. 1985. ``Contiguity and Military Escalation in Major Power
Rivalries, 1816-1980.'' \emph{Journal of Politics} 47(4), 1203-1211.

Douglass, Rex W., Thomas Leo Scherer, J. Andres Gannon, Erik Gartzke.
forthcoming. ``ICBeLLM: High Quality International Events Data with Open
Source Large Language Models on Consumer Hardware.'' In Kyle Beardsley,
Patrick James, and Jonathan Wilkenfeld (eds.), \emph{A Century of
International Crises, 1918-2019}. Ann Arbor: University of Michigan
Press.

Douglass, Rex W., Thomas Leo Scherer, J. Andrés Gannon, Erik Gartzke,
Jon Lindsay, Shannon Carcelli, Jonathan Wilkenfeld, et al.~2022.
``Introducing the ICBe Dataset:~Very High Recall and Precision Event
Extraction from Narratives about International Crises.''
arXiv:2202.07081 {[}Cs, Stat{]},
February.~\href{https://urldefense.com/v3/__https:/arxiv.org/abs/2202.07081__;!!Mih3wA!CLsJAkIaG18LhmH8jhnNw4GVB0of31s_n91IbLKW_lSvbtP4jy3sO8Kgd4zgE_R38RC2ALcN0AsDgo9rF1LIew$}{https://arxiv.org/abs/2202.07081}.

Fearon, James D. 1995. ``Rationalist Explanations for War.''
\emph{International Organization} 49(3): 379---414.

Fearon, James D. 1994. ``Signaling Versus the Balance of Power and
Interests: An Empirical Test of the Crisis Bargaining Model.''
\emph{Journal of Conflict Resolution} 38(2): 236---269.

Fearon, James. 2002. ``Selection effects and deterrence.''
\emph{International Interactions} 28(1): 5-29.

Fey, Mark, and Kristopher W. Ramsay. 2011. ``Uncertainty and Incentives
in Crisis Bargaining: Game‐Free Analysis of International Conflict.''
\emph{American Journal of Political Science}, 55(1), 149-169.

Freedman, Lawrence. 2004. \emph{Deterrence}. Washington DC: Polity
Press.

Freedman, Lawrence. 1981. \emph{The Evolution of Nuclear Strategy}.
London: Palgrave Macmillan.~~

Glaser, Charles L. 1992. ``Political Consequences and Military Strategy:
Expanding and Refining the Spiral and Deterrence Models.'' \emph{World
Politics} 44(4):497---538.

Gannon, J Andrés. 2022. ``One If by Land, and Two If by Sea:
Cross-Domain Contests and the Escalation of International Crises.''
\emph{International Studies Quarterly} 66(4).

Gartzke, Erik. 1999. ``War Is in the Error Term.'' \emph{International
Organization} 53(3): 567--87.

Gartzke, Erik, and Joseph J. Hewitt. 2010. ``International Crises and
the Capitalist Peace.'' \emph{International Interactions}, 36(2),
115-145.

Gelpi, Christopher, and Michael Griesdorf. 2001. ``Winners or Losers?
Democracies in International Crises.'' \emph{American Political Science
Review}, 95(3), 633-47.

Goldstein, Joshua S. 1992. ``A Conflict---Cooperation Scale for WEIS
Events Data.'' \emph{Journal of Conflict Resolution} 36(2): 369---385.

Guolin Ke, Qi Meng, Thomas Finley, Taifeng Wang, Wei Chen, Weidong Ma,
Qiwei Ye, Tie-Yan Liu. 2017. ``LightGBM: A Highly Efficient Gradient
Boosting Decision Tree''. \emph{Advances in Neural Information
Processing Systems} 30:3149-3157.

Heckman, James J., and Burton Singer. 2017. ``Abducting Economics.''
\emph{American Economic Review}, 107 (5): 298-302.

Hegre, Haavard, Nils W. Metternich and Julian Wucherpfennig. 2017.
``Introduction: Forecasting in Peace Research.'' \emph{Journal of Peace
Research}, 54(2), 113-124.

Herz, John. 1950. ``Idealist Internationalism and the Security
Dilemma.'' \emph{World Politics} 2(2):\\
157-180.

Hobbes, T. (1651). \emph{Leviathan or the Matter, Forme, and Power of a
Common-Wealth Ecclesiastical and Civill}. Project Gutenberg Ebook:~
https://www.gutenberg.org/files/3207/3207-h/3207-h.htm.

Huth, Paul K. 1999. ``Deterrence and International Conflict: Empirical
Findings and Theoretical Debates.'' \emph{Annual Review of Political
Science} 2: 25---48.~

Huth, Paul K. 1988. \emph{Extended Deterrence and the Prevention of
War}. Yale University Press.

Huth, Paul and Todd L. Allee. 2002. ``Domestic Political Accountability
and the Escalation and Settlement of International Disputes.''
\emph{Journal of Conflict Resolution} 46(6): 754-790.

Huth, Paul K, Christopher Gelpi, and D. Scott Bennett. 1993. ``The
Escalation of Great Power Militarized Disputes: Testing Rational
Deterrence Theory and Structural Realism.'' \emph{American Political
Science Review}, 87(3), 609-623.

Huth, Paul K., and Bruce Russett. 1988. ``Deterrence Failure and Crisis
Escalation.'' \emph{International Studies Quarterly} 32(1): 29---45.

Huth, Paul K., and Bruce Russett. 1984. ``What Makes Deterrence Work?:
Cases from 1900 to 1980.'' \emph{World Politics} 36(4): 496---526.

Kinsella, David, and Bruce Russett. 2002. ``Conflict Emergence and
Escalation in Interactive International Dyads.'' \emph{The Journal of
Politics}, 64(4), 1045-1068.

Jervis, Robert, Richard Ned Lebow and Janice Gross Stein. 1989.
\emph{Psychology and Deterrence}. Baltimore, MD: Johns Hopkins
University Press.

Jervis, Robert. 1976. \emph{Perception and Misperception in
International Politics}. Princeton, NJ:~ Princeton University Press.

Kahn, Herman. 1965. \emph{On Escalation: Metaphors and Scenarios}. New
York: Praeger.

Kahn, Herman. 1960. \emph{On Thermonuclear War}. Princeton, NJ:
Princeton University Press.

Kissinger, Henry. 1957. \emph{Nuclear Weapons and Foreign Policy}. New
York: Harper.

Kreps, Sarah, and Jacquelyn Schneider. 2019. ``Escalation Firebreaks in
the Cyber, Conventional, and Nuclear Domains: Moving beyond
Effects-Based Logics.'' \emph{Journal of Cybersecurity} 5(1): 1--11.

Kydd, Andrew. 1997. ``Game Theory and the Spiral Model.'' \emph{World
Politics} 49(3):371---383.

Lebow, Richard Ned and Janice Gross Stein. 1989. ``Rational Deterrence
Theory: I Think, Therefore I Deter.'' \emph{World Politics} 41(2):
208---224.

Lin-Greenberg, Erik. 2023. ``Evaluating Escalation: Conceptualizing
Escalation in an Era of Emerging Military Technologies.'' \emph{The
Journal of Politics}. 85(3):1151-1155.

Maoz, Zeev, Paul L. Johnson, Jasper Kaplan, Fiona Ogunkoya and Aaron P.
Shreve. 2019. ``The Dyadic Militarized Interstate Disputes (MIDs)
Dataset, Version 3: Logic, Characteristics and Comparisons to
Alternative Datasets.'' \emph{Journal of Conflict Resolution} 63(3):
811---835.~

McClelland, Charles. 1978. ``World Event Interaction Survey.''
\emph{Inter-university Consortium for Political and Social Research
(ICPSR)} Codebook \#5211. University of Michigan.

Mearsheimer, John J. 1983. \emph{Conventional Deterrence}. Cornell
University Press.

Merrett, Richard L., Robert G. Muncaster and Dina A. Zinnes. 1993.
\emph{International Event-Data Developments: DDIR Phase II}.~Ann Arbor,
MI: University of Michigan Press.

Morgan, Patrick M. 1977. \emph{Deterrence: A Conceptual Analysis}. Los
Angeles, CA: Sage.

Morgenthau, Hans J. 1964. ``The Four Paradoxes of Nuclear Strategy.''
\emph{American Political Science Review} 58(1): 23---35.

Morrow, James D. 2019. ``International Law and the Common Knowledge
Requirements of Cross-Domain Deterrence.'' In \emph{Cross-Domain
Deterrence: Strategy in an Era of Complexity}, edited by Jon R. Lindsay
and Erik A. Gartzke, 1st edition., 187--204. New York, NY: Oxford
University Press.

Moul, William 1988. ``Balance of Power and the Escalation to War of
Serious Disputes among the European Great Powers, 1815-1939: Some
Evidence.'' \emph{American Journal of Political Science} 32(2), 241-275.

Nalebuff, Barry. 1986. ``Brinkmanship and Nuclear Deterrence: The
Neutrality of Escalation.'' \emph{Conflict Management and Peace Science}
9(1), 19-30.

O'Neill, Barry. 1991. ``Conflictual Moves in Bargaining: Warnings,
Threats, Escalations, and Ultimatums.'' In \emph{Negotiation Analysis},
edited by H. Peyton Young, 87--108. University of Michigan Press.

Powell, Robert. 1989. ``Crisis Stability in the Nuclear Age.''
\emph{American Political Science Review} 83(1): 61--76.

Powell, Robert. 1990. \emph{Nuclear Deterrence Theory: The Search for
Credibility}. Cambridge, UK:~ Cambridge University Press.

Quackenbush, Stephen 2011. \emph{Understanding General Deterrence:
Theory and Application}. New York:~ Palgrave Macmillan.~

Quackenbush, Stephen 2010. ``General Deterrence and International
Conflict: Testing Perfect Deterrence Theory.'' \emph{International
Interactions} 36(1): 60-85.

Raleigh, Clionadh, Andrew Linke, Haavard Hegre and Joakim Karlsen. 2010.
``Introducing ACLED: An Armed Conflict Location and Events Dataset.''
\emph{Journal of Peace Research} 47(5): 652---660.

Reed, William. 2000. ``A Unified Statistical Model of Conflict Onset and
Escalation.'' \emph{American Journal of Political Science} 44(1), 84-93.

Rovner, Joshua. 2020. ``Give Instability a Chance?'' \emph{War on the
Rocks}, July 28.
https://warontherocks.com/2020/07/give-instability-a-chance/.

Sagan, Scott D. 1993. \emph{The Limits of Safety: Organizations,
Accidents and Nuclear Weapons}. Princeton, NJ: Princeton University
Press.

Sagan, Scott D. 1989. \emph{Moving Targets: Nuclear Strategy and Nuclear
Security}. Princeton, NJ: Princeton University Press.

Schelling, Thomas. 1966. \emph{Arms and Influence}. New Haven, CT: Yale
University Press.

Schrodt, Philip A. 2007. ``Inductive event data scaling using item
response theory.'' In \emph{Summer Meeting of the Society of Political
Methodology}. Available at http://eventdata.psu.edu.

Schrodt, Phillip A. 2012. ``CAMEO: Conflict and Mediation Event
Observations Event and Actor Codebook.'' Typescript. The Pennsylvania
State University.

Schultz, Kenneth A. 1998. ``Domestic Opposition and Signaling in
International Crises.'' \emph{American Political Science Review}, 92(4),
829-844.

Sechser, Todd S., and Matthew Fuhrmann. 2013. ``Crisis Bargaining and
Nuclear Blackmail.'' \emph{International Organization} 67(1):
173---195.~

Sherman, Frank. 1994.~ ``SHERFACS: A Cross-Paradigm, Hierarchical and
Contextually Sensitive Management Dataset.'' \emph{International
Interactions} 20(1): 79---100.

Shirkey, Zachary. 2016. ``Uncertainty and War Duration.''
\emph{International Studies Review} 18(2): 244---267.

Shirkey, Zachary. 2012. ``When and How Many: The Effects of Third Party
Joining on Casualties and Duration in Interstate Wars.'' \emph{Journal
of Peace Research} 49(2): 321-334.~

Slantchev, Branislav L. 2005. ``Military Coercion in Interstate
Crises.'' \emph{American Political Science Review} 99(4): 533-547

Smith, Alastair. 1999. ``Testing Theories of Strategic Choice: The
Example of Crisis Escalation.'' \emph{American Journal of Political
Science} 43(4): 1254---1283.~

Smith, Alastair. 1998. ``International Crises and Domestic Politics.''
\emph{American Political Science Review}, 92(3), 623-638.

Smith, Alastair. 1995. ``Alliance Formation and War.''
\emph{International Studies Quarterly} 39(4): 405---425.

Snyder, Glenn H. 1961. \emph{Deterrence and Defense}. Princeton, NJ:
Princeton University Press.

Snyder, Glenn H., and Paul Diesing. 1977. \emph{Conflict Among Nations:
Bargaining, Decision Making and System Structure in International
Crises}. Princeton, NJ: Princeton University Press.

Sorokin, Gerald L. 1994. ``Alliance Formation and General Deterrence:~A
Game-Theoretic Analysis.'' \emph{Journal of Conflict Resolution} 38(2):
298---325.

Spaniel, William, and Işıl İdrisoğlu. 2023. ``Endogenous Military
Strategy and Crisis Bargaining.'' \emph{Conflict Management and Peace
Science}, 0(0).

Stam, Allan. 1996. \emph{Win, Lose or Draw: Domestic Politics and the
Crucible of War}. University of Michigan Press.

Tavory, Iddo, and Timmermans, Stefan. 2014. \emph{Abductive Analysis:
Theorizing Qualitative Research}. United Kingdom, University of Chicago
Press.

Wagner, R. Harrison. 2007. \emph{War and the State}. Ann Arbor, MI:
University of Michigan Press.

Wagner, R. Harrison. 2000. ``Bargaining and War.'' \emph{American
Journal of Political Science} 44(3):469---484.~

Waltz, Kenneth 1990. ``Nuclear Myths and Political Realities.''
\emph{American Political Science Review} 84(3): 731---745.~

Weisiger, Alex. 2013. \emph{Logics of War}. Ithaca, NY: Cornell
University Press.~

Wohlsletter, Albert. 1959. ``The Delicate Balance of Terror.''
\emph{Foreign Affairs} 37: 209---234.

Zagare, Frank C. and D. Marc Kilgour. 2000. \emph{Perfect Deterrence}.
Cambridge, UK:~ Cambridge University Press.

Zagare, Frank C. and D. Marc Kilgour. 1998. ``Deterrence Theory and the
Spiral Model Revisited.'' \emph{Journal of Theoretical Politics}
10(1):59---87.

\end{document}